\documentclass[prl,aps,twocolumn,epsf]{revtex4}
\usepackage{epsf,amsmath,graphicx,verbatim}
\usepackage{color}
\usepackage{amsmath}
\usepackage{amssymb}
\begin{document}
\textheight=26.0truecm
\textwidth=16truecm
\renewcommand{\rmdefault}{ptm}
\oddsidemargin=0.0cm
\topmargin=-1.0cm
\headsep=0mm
\raggedbottom
\pagestyle{empty}
\unitlength=1cm

\title{Entropy driven formation of Smectic C in system of Zig-zag shaped molecules}
\author{Prabal K. Maiti$^{1,*}$, Yves Lansac$^{1,!}$, Matthew A. Glaser and Noel A. Clark}
\affiliation{Ferroelectric Liquid Crystal Material Research Center,\\
Department of Physics, University of Colorado, Boulder, CO 80309}
\begin{abstract}
We have carried out Monte Carlo simulations of zig-zag shaped molecules where the
the molecules is composed of three rigidly linked hard spherocylinder arranged 
in a zigzag fashion. By varying zigzag angle we have mapped out the whole phase
diagram as a function of pressure, molecular density and zigzag angle $\Psi$. For 
$\Psi$ between $35^0$ and $80^0$ our model simulation exhibits SmC phase. This is
the first conclusive evidence where steric interaction arising out of molecular shape alone 
induce the occurrence of SmC phase for a wide range of zigzag angle. For smaller 
$\Psi$, a transition from tilted crystal (XT) to crystal is observed.
\end{abstract}
\maketitle
Liquid crystal phase is very sensitive to the molecular shape. Relating the macroscopic
properties of liquid crystal phase to the microscopic structures of the molecules is 
very complex problem. On the other hand it is essential to understand this 
structure-properties relationship both from fundamental point of view as well as
for different technological applications. Theoretically
it is a very difficult task since it involves hundreds of molecules and their mutual
interactions. Computer simulations has emerged as an important tool to investigate 
the dependence of the liquid crystal phase behavior to the structure of the constituent 
molecules. Various models has been used such as simple spherocylinder or
hard-sphere chains which interact through
hard or soft excluded-volume repulsion \cite{macgroth,frenkel,dave}, molecules with
ellipsoid shape interacting through Gay-Berne potential \cite{gay,bates}, 
simple ``bead-spring" representation of molecules interacting through 
Lennard-Jones potential \cite{affo}. There are also simulations involving atomic level models
of real liquid crystal molecules \cite{wilson, matt}.

There are a number of theoretical works which gives a microscopic theories for the 
formation of smectic C phase \cite{mcmillan,cabib,meer,gooss,velasco,wulf}.
In McMillan theory it is the electric dipole-dipole interactions which 
produce tilt in the smectic C. Later quadrupole-quadrupole interaction was 
introduced as a perturbation of smectic A phase to get a smectic A to smectic C transition
\cite{gooss}. 
Motivated by the large discrepancy observed between X-ray and the optical data
for the molecular tilt angle in smectic C materials, Bartolino {\sl et al.} \cite{bartolino}, 
resuming an idea from Guillon and Skoulios \cite{guillon}, proposed
a molecular model (known as the zigzag model) derived from lyotropic systems: 
the rigid central core, optically anisotropic, imposes the tilt of the optical
axis and the melted aliphatic end chains are on average closer to the normal 
to the layers. 
Wulf \cite{wulf} while studying the zigzag shaped molecules concluded that
the tilted smectic C is formed due to the steric interactions arising out of the
packing arrangement of such molecules. 
There are number of simulations concerning the occurrence of smectic C phase, but 
none of them are conclusive. Neal and Parker has performed simulations of  model
molecules comprising three rigidly linked Gay-Berne (GB) sites arranged in a zigzag
fashion \cite{neal} and but did not find the occurrence of smectic C phase. Evidence 
of smectic C phase with varying tilt angle from layer to layer 
was reported in simulation of zigzag shaped molecules comprising 
seven rigidly bonded soft spheres \cite{Xu}.
One of the major goal of this paper is to investigate if steric interaction arising 
from molecular shape alone can induce tilt in the smectic phase.

The model molecule is comprised of three rigidly linked hard 
spherocylinders of length/breadth ratio $L_{\rm rod}/D$ arranged in a zigzag configuration
(see the inset in figure 1). Both ends of the molecule is making an angle 
$\Psi$ with the core. The idea behind using such
hard core model is that liquid crystal phase behavior is largely
entropy driven and determined by the hard core repulsion between the liquid
crystal mesogens. Hard spherocylinder provides a simple model both in terms
of computational ease and theoretical approach. It has been studied extensively 
and exhibit a rich phase behavior including isotropic, nematic, smectic A, 
and crystal phases. The advantage of using such hard core model is that we can 
vary the shape of the molecule (by changing $\Psi$) and see how that affects 
the large-scale organization of the liquid crystal phases. 

For convenience we introduce reduced units. The reduced pressure $P^*$ is
defined as $P^* = \beta Pv_{hsc}$ and a reduced density $\rho^* = \rho v_{hsc}$,
where $v_{hsc}$ is the volume of the straight hard spherocylinder of
length $L$ and breadth $D$.
We have performed MC simulation in $NPT$ ensemble
with periodic boundary condition on a system of $400$ zig-zag shaped molecules.
The simulation cell consists
of $N = 400$ molecules in a cubic box of dimension $L_x \times L_y \times L_z$. 
Initially the system was prepared in a crystal phase at high pressure.
Among the possible different crystalline order, we choose the antipolar 
crystal ordering corresponding to the highest packing density to ensure
the highest stability. Starting from crystal phase at high pressure ($P^* = 13$)
we decrease the pressure successively by steps of $\Delta P^{\star} = 1$,
until we reach a reduced pressure $P^{\star} = 1$.
For each run, at a
given pressure, the final equilibrated configuration obtained from the
previous higher pressure is used as the initial state. At each state point
($P^\star$, $\Psi$) the system is equilibrated for $200000$ MC cycles and
one million MC cycles are used for the production of the results and the
analysis of the various thermodynamical and structural quantities.
During each MC step each molecules were chosen
randomly and displaced using Metropolis criteria. The reorientation move was
performed using quternion \cite{allen}. In all the simulations 
reported below, we use a length to breadth ratio $L_{\rm ban}/D = 2.$

In order to fully characterize different phases of the systems various 
order parameters were computed.
The location of the solid--liquid phase boundary is determined
by computing the squared-magnitude of the in-layer translation order parameters
${{\rho}_{\bf G}}_{k}$ defined as
\begin{equation}
{{\rho}_{\bf G}}_{k} = \frac{1}{M} \sum_{j = 1}^{M} \exp ({i {\bf G}_{k}
\cdot {\bf r}_{j}})
\end{equation}

${\bf G}_{1}, {\bf G}_{2}, {\bf G}_{3}$ are
the reciprocal basis vectors and ${\bf r}_{j}$ is the position of the center
of mass of the molecule $j$ and $M$ is the number of molecules in a given
layer.

The smectic--nematic phase boundary is determined by the squared-magnitude
of the layer translational order parameter ${\rho}_{\parallel}$ defined as
\begin{equation}
{\rho}_{\parallel} = \frac{1}{N} \sum_{j = 1}^{N} \exp ({i {\bf G}_{\parallel}
\cdot {\bf r}_{j}})
\end{equation}

where
${\bf G}_{\parallel} = \frac{2 \pi}{d} \; \hat{{\bf z}}$, the layer normal
being along $\hat{{\bf z}}$.

To distinguish between a tilted and a non tilted phase 
(smectic or crystal) we introduce the in-layer polar order parameter 
$\hat{{\bf m}}$ defined as 
\begin{equation}
\hat{{\bf m}} = \frac{1}{M} \sum_{j=1}^{M} \hat{{\bf m}}_{j}
\end{equation}
where
$\hat{{\bf m}}_{j}$ is the unit vector contained in the plane of the
molecule and passing through one of the apex of the molecule. 
With this definition, care must be taken to ensure that two zigzag molecules 
having the z-component of the end-to-end vector pointing in opposite direction 
have in fact the same polar direction.

The orientational order--isotropic phase
boundary is determined by the eigenvalues of the second-rank tensorial
orientational order parameter $Q_{\alpha \beta}$ defined as
\begin{equation}
Q_{\alpha \beta} = \frac{1}{N} \sum_{j=1}^{N} \left({ \frac{3}{2}
n_{i_{\alpha}} n_{j_{\beta}} - \frac{1}{2} {\delta}_{\alpha \beta} }\right)
\end{equation}
with $\alpha, \beta = x,\; y, \; z$
and ${\bf n}_{j}$ is the molecular end-to-end unit vector of molecule $j$.
The nematic order parameter S is given by the largest eigenvalue of the ordering
tensor $Q_{\alpha \beta}$. The value of S is close to zero in the isotropic 
phase and will tend to one in highly ordered phase.

To distinguish between SmA and SmC we compute the tilt angle of the central 
spherocylinder (core) of the zigzag shaped molecule with the layer normal. 

\begin{figure}[!tbp]
\includegraphics[width=3.2in]{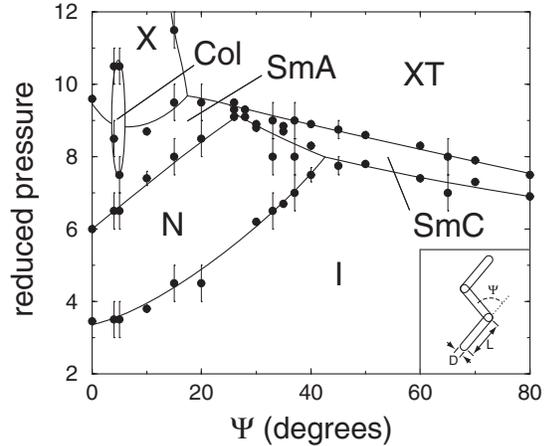}
\caption{\protect{Phase diagram of zigzag shape molecules with aspect ratio L/D = 2
as a function of opening angle $\Psi$ and reduced pressure P*.
The following phase are present: isotropic liquid (I), nematic (N),
smectic A (SmA), smectic C (SmC), columnar (Col), tilted crystal (XT),
crystal (X).}}
\label{phase_diaP}
\end{figure}

To map out the complete phase diagram we have performed simulations for several 
zig-zag angles in between $\Psi = 0$ and $80^0$. The phase diagram 
$(\Psi, P^*)$ is presented in Fig.~\ref{phase_diaP}.

\begin{figure}[!tbp]
\includegraphics[width=3.0in]{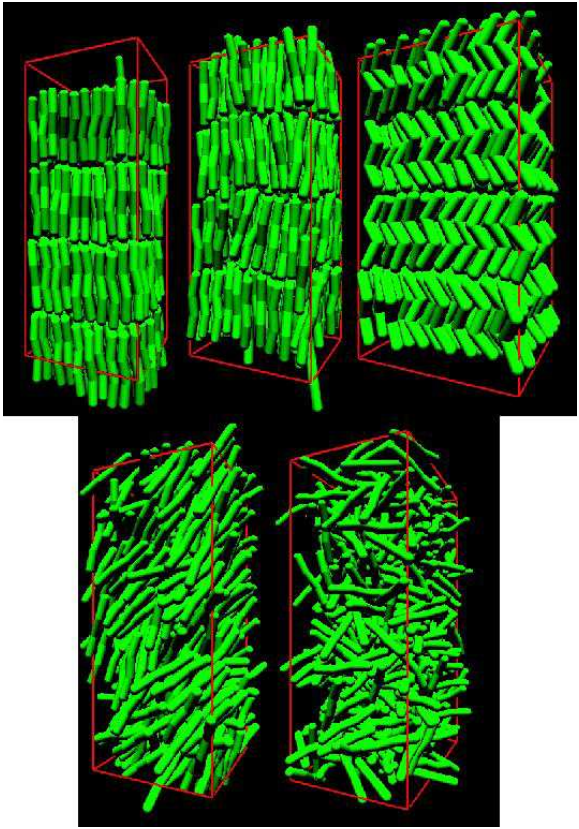}
\caption{\protect{Final configurations from Monte Carlo simulations of N = 400
zigzag molecules as a function of pressure. From
left to right, top to bottom: crystal ($\Psi = 15^{\circ}$, $P^{\star} = 11$),
smectic A ($\Psi = 15^{\circ}$, $P^{\star} = 9$),
smectic C ($\Psi = 65^{\circ}$, $P^{\star} = 8$),
nematic phase ($\Psi = 15^{\circ}$, $P^{\ast} = 5$),
isotropic phase ($\Psi = 15^{\circ}$, $P^{\ast} = 1$).}}
\label{snapshot2}
\end{figure}

The system exhibits rich phase behavior with isotropic liquid (I), nematic (N),
smectic A (SmA), smectic C (SmC), columnar (Col), tilted crystal (XT),
crystal (X) phases. Configurations from the isotropic, nematic, smectic A and
tilted crystal phases are shown in Fig.~\ref{snapshot2} for a zigzag angle 
$\Psi = 15^\circ$. Also shown on the same figure a configuration from the
smectic C phase for a zigzag angle $\Psi = 65^\circ$. 
The phase diagram is completely isomorphous
to the phase diagram obtained for bent-core molecules represented as 
hard-core dimer formed by two interdigitated hard-core spherocylinders
sharing one spherical end cap \cite{lansac}. 
The SmC replaces the polar SmA (SmAP) phase found in bent-core molecules since
polar symmetry breaking leads, for the zigzag model, to a phase having all 
the symmetries defining a tilted smectic phase. 
Close-packing interactions combined to the molecular zigzag geometrical shape 
induce a tilt of the molecules with respect to the layer normal. 

The nematic phase is stable  for zigzag angles smaller than
$\Psi$ = 40$^{\circ}$.
With increasing zigzag angle, the region of stability of the nematic
phase decreases, vanishing for zigzag
angles larger than $\sim 40 ^{\circ}$, leading to an (I, N, SmC) triple
point near $\Psi = 40 ^{\circ}$. The existence of a biaxial nematic phase 
remains an elusive possibility in thermotropic LCs. Zigzag molecules 
are good candidate to exhibit biaxial nematic due to their geometrical 
phase. However, no such phase have been found in our study, confirming 
the results obtained with bent-core molecules \cite{lansac}. As is clear
from simulations of the hard biaxial ellipsoid system \cite{allen90, camp97}, a biaxial nematic
phase requires a highly biaxial molecular shape. Because such a requirement 
could be satisfied for larger L/D and specific zigzag angle in the present 
model, the possibility of the presence of a biaxial phase should not be 
ruled out. 

Because straight spherocylinders do not exhibit any SmC ordering,
it is expected that our model should exhibit a transition from 
SmA to SmC. This transition occurs for a zigzag 
angle between $26 ^{\circ}$ and $28 ^{\circ}$, and is associated with two
triple points, a (SmA, SmC, N) triple point near $\Psi = 28^{\circ}$
and a (SmA, SmC, XT) triple point near $\Psi = 26^ \circ$.
The SmA--SmC transition is purely entropy driven and arise from packing effect. 
In figure \ref{tilt} we have plotted 
the smectic tilt angle (defined as the average over the tilt angle with respect
to the layer normal  made by 
the central spherocylinder of every molecules) as a function of the zigzag 
angle. This is the first time that an idealized model system gives conclusive 
evidence of the occurrence of a SmC phase.
Due to the weak coupling between adjacent layers in the SmC phase, it was 
impossible to determine the relative stability of synclinic (a uniform tilt
direction in all layers) and anticlinic order by direct simulations. 
But it is reasonable to assume that the zigzag model obeys the
same general thermodynamic mechanism like the hard spherocylinder and the
hard-dimer models that we have recently studied \cite{glaser,lansac}. In these 
studies within a framework of "sawtooth" model, we have demonstrated that 
the entropy content of molecular-scale fluctuations of 
the interface between smectic C layers ("out-of-layer" molecular fluctuations) 
provides a general thermodynamic mechanism that uniquely favors synclinic 
ordering.

\begin{figure}[!tbp]
\includegraphics[width=3.3in]{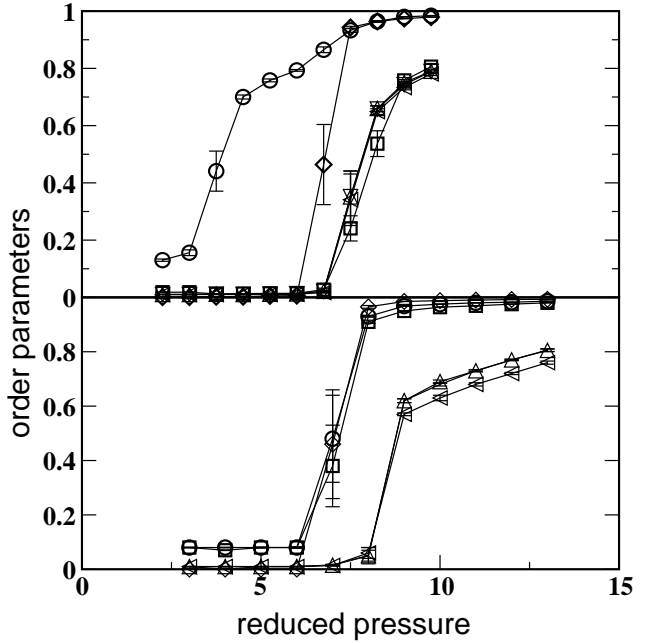}
\caption{\protect{Evolution of the squared-magnitude of order parameters as a function of reduced
pressure for an opening angles of $\Psi = 20 ^\circ$ (top) and
$\Psi = 65 ^\circ$ (bottom) showing, respectively
the phase sequences XT--SmA--N--I and XT--SmC--I as a function
of decreasing pressure.
The following order parameters are plotted: ($\vartriangle$, $\triangledown$,
$\vartriangleleft$)
solid-liquid order parameters ${\rho}_{1}$, ${\rho}_{2}$, ${\rho}_{3}$;
($\square$) polar order parameter
${\bf m}$; ({\Large$\diamond$}) smectic order
parameter $\rho_{\parallel}$; ({\Large$\circ$}) the largest eigenvalue
of the nematic order parameter $Q_{\alpha \beta}$.}}
\label{order_parameter}
\end{figure}

\begin{figure}[!tbp]
\includegraphics[width=3.3in]{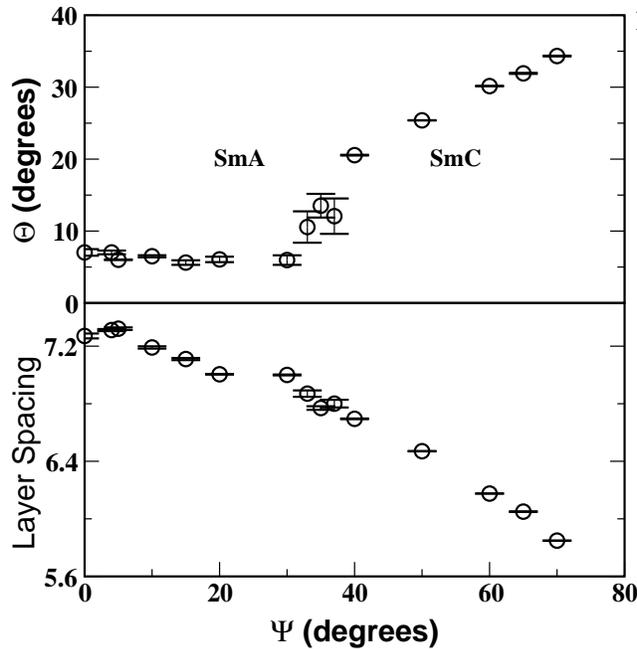}
\caption{\protect{Tilt angle ($\theta$) (top) and layer spacing (bottom) as a 
function of zigzag angle.}}
\label{tilt}
\end{figure}

We also find a transition between a tilted crystalline phase and a
untilted crystalline phase (i.e a rotator phase).
This rotator phase is stable for zigzag angles smaller than $\Psi =
$ 20$^{\circ}$,
and is characterized by a (SmA, X, XT) triple point around
$\Psi =$ 18$^{\circ}$.
Quite interestingly, the rotator phase competes with a columnar phase for
zigzag angles larger than $\Psi \sim $ 3$^{\circ}$ and smaller than
$\Psi \sim$6$^{\circ}$. This narrow columnar phase is characterized by
significant two-dimensional crystal order parameters but a negligible 
magnitude of the smectic order parameter.

Insights into the shape of the phase boundaries can be gained by supposing,
to a first approximation, that the partition function of the system can be
decomposed into a product of positional and orientational contributions,
in which case the entropy is the sum of an orientational entropy
and a translational entropy.
Competition between different forms of entropy
determines the stability of a given phase at a given density. In the limit
of straight spherocylinders ($\Psi = 0^{\circ}$), the isotropic--nematic phase transition occurs
when the gain in positional entropy $S^{\rm pos}$ exceeds the loss of
orientational entropy $S^{\rm orient}$ \cite{onsager}.
A nematic--smectic phase transition
occurs when the gain in translational entropy perpendicular to the long
molecular axis $S^{\rm pos}_{\perp}$ exceeds the loss of positional
entropy parallel to the long molecular axis $S^{\rm pos}_{\parallel}$, leading
to the formation of a stack of two-dimensional liquid layers.
Similar reasoning can be applied to zig-zag molecules: in the range
$0 ^{\circ} < \Psi < 40 ^{\circ}$, the isotropic phase is more favourable
at smaller zigzag angles. As the cores become more bent (larger zigzag angles),
the gain in positional entropy associated with nematic ordering is reduced,
and the nematic phase range is reduced, eventually disappearing for
$\Psi > 40^{\circ}$.
The shape of the nematic--SmC boundary (i.e., for $30^{\circ} < \Psi <
40^{\circ}$)
can be qualitatively understood in the same way by noticing that the
positional entropy parallel to the long molecular axis
$S^{\rm pos}_{\parallel}$ is larger for smaller zigzag angles than for
larger ones, stabilizing the nematic phase for smaller
zigzag angles. 

The model exhibits a rich phase behavior including tilted and nontilted
crystal, columnar, smectic A, smectic C, nematic and isotropic phases. The
model shows without any ambiguity that excluded volume interaction arising 
out of molecular shape is sufficient to produce  tilted smectic phase even
in the absence of electrostatic interactions.

This work was supported by NSF MRSEC Grant DMR 98-09555.


\begin{references}
\item[$^{1}$] Present address: Materials and Process Simulation Center, 139-74
California Institute of Technology, Pasadena, CA 91125, U.S.A.
\item[$*$] Email : maiti@wag.caltech.edu.
\item[$!$] Email : lansac@wag.caltech.edu.
\end{references}

\begin{thebibliography}{99}
\bibitem{degens} P. G. de Gennes and J. Prost, in {\sl The Physics of
Liquid Crystals}, 2nd ed. (Oxford University Press, New York, 1993).

\bibitem{macgroth} S. C. McGrother, D. C. Williamson, and G. Jackson,
J. Chem. Phys. {\bf104}, 6755 (1996).

\bibitem{frenkel} P. Bolhuis and D. Frenkel, J. Chem. Phys. {\bf106}, 666 (1997).

\bibitem{dave} D. C. Williamson and G. Jackson, J. Chem. Phys., {\bf108}, 10294 (1998).
\bibitem{gay} J. G. Gay and B. J. Berne, J. Chem. Phys. {\bf74}, 3316 (1981).
\bibitem{bates} M. A. Bates and G. R. Luckhurst, J. Chem. Phys., {\bf110}, 7087 (1999).
\bibitem{wilson} M. R. Wilson and M. P. Allen, {\sl Mol. Cryst. Liq. Cryst.}, {\bf198}, 
465 (1991)

\bibitem{matt} M. A. Glaser, R. Malzbender, N. A. Clark, and D. M. Walba, {\sl J. Phys.:
Condens. Matter}, {\bf6}, A261 (1994).

\bibitem{glaser} M. A. Glaser and N. A. Clark, {\sl Phy. Rev. E.}, {\bf66}, 021711 (2002).

\bibitem{affo} F. Affouard, M. Kr\"oger and S. Hess, Phys. Rev. E., {\bf54}, 5178 (1996).

\bibitem{Xu} J. Xu, R. L. B. Selinger, J. V. Selinger, B. R. Ratna and R. Shashidar,
Phys. Rev. E., {\bf60}, 5584 (1999).


\bibitem{neal} M. P. Neal, A. J. Parker and C. M. Care, Mol. Phys., {\bf91}, 603 (1997).

\bibitem{bartolino} R. Bartolino, J. Doucet and G. Durand, {\sl Ann. Phys.}, {\bf3}, 389 (1978)
\bibitem{guillon} D. Guillon and A. Skoulios, {\sl J. Physique (Paris)}, {\bf38}, 79 (1977).
\bibitem{Ian} I. M. Withers, C. M. Care and D. J. Cleaver, {\bf113}, 5078 (2000).

\bibitem{Xu1} J. Xu, R. L. B. Selinger, J. V. Selinger and R. Shashidar (preprint).
\bibitem{mcmillan} W. L. McMillan, Phys. Rev. A{\bf11}, 365 (1974).
\bibitem{cabib} D. Cabib and L. Benguigui, J. Phys. (Paris) {\bf38}, 419 (1977).
\bibitem{meer} B. W. van der Meer and G. Vertogen, J. Phys. (Paris), Colloq. {\bf40},
C3-222 (1979).
\bibitem{gooss} W. J. A. Goossens, Europhys. Lett., {\bf3}, 341 (987).
\bibitem{velasco} E. Velasco, L. Mederos, and T. J. Sluckin, Liq. Cryst., {\bf20},
399 (1996).
\bibitem{wulf} A. Wulf, Phys. Rev. A{\bf11}, 365 (1975).
\bibitem{allen} M. P. Allen and D. J. Tildesley, {\sl Computer Simulation of
Liquids}, (Oxford, 1987).
\bibitem{lansac} Y. Lansac, P. K. Maiti, N. A. Clark, and M. A. Glaser, {\sl Phys.
Rev. E.}, {\bf67}, 011703 (2003).
\bibitem{allen90} M.~P.~Allen,
{\sl Liq.~Cryst.}~{\bf 8}, 499 (1990).

\bibitem{camp97} P.~J.~Camp, M.~P.~Allen,
{\sl J.~Chem.~Phys.}~{\bf 106}, 6681 (1997).
\bibitem{onsager} L.~Onsager, {\sl Ann.~(N.~Y.)~Acad.~Sci.}~{\bf 51}, 627
(1949).
\end{thebibliography}
\end{document}